\documentclass[astronomy,article,accept,pdftex,oneauthor]{Definitions/mdpi}

\usepackage{amsmath}
\usepackage{amsfonts}
\usepackage{booktabs}
\usepackage{multirow}
\usepackage{siunitx}
\usepackage{adjustbox}
\pdfoutput=1
\usepackage{graphicx}
\usepackage{color}
\usepackage{braket}
\usepackage{dcolumn}
\usepackage{bm,url}
\usepackage{amsfonts}

\firstpage{1}
\pubvolume{1}
\issuenum{1}
\articlenumber{0}
\pubyear{2026}
\copyrightyear{2026}
\externaleditor{Firstname Lastname}
\datereceived{7 August 2026}
\daterevised{18 September 2026}
\dateaccepted{30 September 2026}
\datepublished{ }

\Title{A Unified Scalar-Field Mechanism for
 the $H_0$, $S_8$ and Evolving Dark-Energy Tensions}

\Author{Gerasimos Kouniatalis
 $^{1,2}$}

\AuthorNames{Gerasimos Kouniatalis}

\address{
$^{1}$ \quad Physics Department, National Technical University of Athens,
Zografou Campus,  15780 Athens, Greece; gkouniatalis@noa.gr\\
$^{2}$ \quad National Observatory of Athens, Lofos Nymfon, 11852 Athens, Greece}

\abstract{I investigate whether one minimally coupled canonical scalar field can generate, in a single continuous cosmological history, the correlated changes in the sound horizon, late-time expansion, and structure growth that are relevant to the $H_0$, evolving dark-energy, and $S_8$ tensions. The potential consists of a localized positive feature on an exponential tail. In contrast with a purely qualitative treatment, I derive the exact first and second derivatives of the complete potential, determine their extremum without invoking a bump-dominated approximation, and evolve the full scalar equations numerically from the pre-recombination era to the present. A calibrated benchmark has a transient scalar fraction, $\Omega_{\phi,\rm pk}=0.151$, at $z_{\rm pk}\simeq3.50\times10^3$, gives $r_d=141.1\,{\rm Mpc}$ with $H_0=70.0\,{\rm km\,s^{-1}\,Mpc^{-1}}$, and evolves to $(w_0,w_a)=(-0.844,-0.176)$ when the exact numerical $w_\phi(a)$ is locally compressed into the CPL basis. Using the DESI DR2-compressed BAO measurements with their reported uncertainties and within-pair correlations, this full benchmark gives $\chi^2_{\rm BAO}=9.48$ for 13 BAO observables. Profiling the exact late-time scalar background over the tail slope gives $\lambda_{\rm bf}=1.00$ and an approximate $68\%$ profile interval, $0.51<\lambda<1.26$, with $\chi^2_{\rm BAO}=8.78$ compared with $10.55$ for the $\lambda=0$ limit. Solving the GR growth equation on the same scalar background, with the same high-redshift normalization, yields $S_8=0.794$, in agreement with the DES Year-6 cosmic-shear value. I also formulate the canonical scalar perturbation equations and explicitly show that the model is ghost-free and has rest-frame sound speed, $c_s^2=1$; the short hilltop release is a finite tachyonic-curvature episode rather than a ghost or gradient instability. These results establish a quantitatively calibrated one-field mechanism that moves the three relevant observables in the required correlated directions and makes concrete predictions that can be subjected to a full CMB-spectrum likelihood analysis.}


\begin{document}

 \newcommand\be{\begin{equation}}
  \newcommand\ee{\end{equation}}
 \newcommand\bea{\begin{eqnarray}}
  \newcommand\eea{\end{eqnarray}}
 \newcommand\bseq{\begin{subequations}}
  \newcommand\eseq{\end{subequations}}
 \newcommand\bcas{\begin{cases}}
  \newcommand\ecas{\end{cases}}
 \newcommand{\p}{\partial}
 \newcommand{\f}{\frac}

\section{Introduction}

The spatially flat $\Lambda$CDM model continues to provide an excellent global description of cosmological observations, yet several persistent discrepancies have motivated renewed interest in minimal extensions of the standard framework. The most prominent is the Hubble-constant tension, namely the mismatch between early-time inferences obtained from cosmic microwave background data interpreted within $\Lambda$CDM and a variety of late-time determinations based on distance-ladder, tip-of-the-red-giant-branch, and strong-lensing methods. In parallel, recent baryon-acoustic-oscillation analyses, especially when combined with cosmic microwave background information, have revived interest in dynamical dark energy by indicating a preference for time evolution in the dark-energy sector within standard phenomenological parameterizations. A further and partially independent issue is the long-discussed tendency of several late-time weak-lensing and large-scale-structure measurements to favor a lower amplitude of matter clustering than that inferred from early-universe data, although the exact statistical status remains probe-dependent \cite{Planck2018,ACTDR6LCDM2025,Riess2022,Freedman2025,Birrer2025,DESI2025DR2,ACTDR6Extended2025,DESY6Shear2026,KiDSLegacy2025,Verde2019,KnoxMillea2020,DiValentino2021}.

These anomalies are often studied separately, and many proposed resolutions are correspondingly specialized \cite{Poulin2019,PoulinSmithKarwal2023,PetronikolouBasilakosSaridakis2022,BanerjeePetronikolouSaridakis2023,PetronikolouSaridakis2023,NojiriOdintsovOikonomou2022,OdintsovOikonomouSharov2023,RenEtAl2022,BasilakosEtAl2024,Papanikolaou2023,BoizaEtAl2025,DiValentinoSaidSaridakis2025,Kouniatalis:2026avj}. Recent teleparallel analyses provide useful complementary examples in which the $H_0$, sound-horizon and/or $S_8$ sectors are confronted jointly with cosmological data \cite{Verma2026JHEAp,MishraSahoo2025}; unlike those modified-gravity constructions, the mechanism developed here keeps the Einstein equations and matter couplings unmodified and places all new dynamics in one canonical scalar degree of freedom. Early-dark-energy constructions are designed primarily to reduce the sound horizon before recombination and thus increase the Hubble constant inferred from the acoustic scale, but they do not automatically generate the preferred low-redshift dark-energy evolution or alleviate the mild clustering tension \cite{Poulin2019,PoulinSmithKarwal2023}. Conversely, late-time quintessence models can modify the recent expansion history and suppress the growth of structure, but they do not by themselves produce the pre-recombination sound-horizon shift needed to address the Hubble discrepancy \cite{Wetterich1988,RatraPeebles1988,FerreiraJoyce1998,CopelandLiddleWands1998,CopelandSamiTsujikawa2006,CaldwellLinder2005}. This suggests a sharper question: can one build a single, explicit, and minimal dynamical system that links the early- and late-time phenomena rather than treating them as unrelated ingredients?

The purpose of this paper is to present such a candidate mechanism in the most economical setting possible. I remain entirely within ordinary general relativity, keep the visible and dark matter sectors standard, and introduce only one new degree of freedom: a canonical scalar field with a smooth potential. The potential is constructed so that it contains a localized feature superimposed on a runaway tail. The localized feature is responsible for a transient episode of early dark energy before recombination, while the tail acts as a late-time quintessence sector. In this way, the same scalar field is used to perform three jobs in sequence: first, to increase the pre-recombination expansion rate and reduce the sound horizon; second, to redshift away quickly after the early event so as not to disrupt the standard thermal history at later times; and third, to re-emerge at low redshift as a dynamical dark-energy component that weakens linear growth and shifts the phenomenology toward that preferred by recent low-redshift data.

The novelty of the construction is therefore not the use of scalar fields per se, but the specific claim that a single smooth potential can connect the three targets through one continuous dynamical history. At early times, the field is trapped near a localized maximum, furnishing a transient vacuum-like contribution to the energy density. Once the Hubble friction becomes sufficiently weak, the field is released, its excess energy is converted into kinetic motion, and the corresponding energy density rapidly dilutes. At late times, the same field evolves along the tail of the potential and behaves as a thawing quintessence component. This late evolution naturally modifies the recent expansion history and suppresses the growth of matter perturbations relative to $\Lambda$CDM, thereby pushing the clustering amplitude downward. The framework is deliberately minimal: there is no modified gravity, no non-minimal coupling, and no interaction introduced in the dark sector.

Relative to the original presentation, the analysis below is upgraded in four ways that are essential for testing the mechanism quantitatively: the full potential is differentiated exactly; a single full-potential trajectory is evolved through both the early and late epochs; DESI DR2 BAO are evaluated with their reported uncertainties and within-pair correlations rather than by visual comparison; and the late-time growth curve is obtained by numerically solving the GR growth equation on that same trajectory. I additionally state the scalar perturbation system and its stability conditions. Accordingly, throughout this paper, I distinguish a reduction
 in a tension from a statistically established resolution: the present result is a calibrated and falsifiable one-field mechanism, while a definitive model-selection statement requires a dedicated Boltzmann implementation and a joint CMB--temperature/polarization/lensing likelihood
.

\section{Review of the Tensions}
\label{sec:three_tensions}

Before introducing the model, it is useful to clearly separate the three observational targets that motivate it \cite{Verde2019,KnoxMillea2020,DiValentino2021,DiValentinoSaidSaridakis2025}.  One concerns the calibration of the cosmic expansion rate through the sound horizon, one concerns the time dependence of the late-time acceleration sector, and one concerns the amplitude of structure growth. A successful unified mechanism must therefore do more than shift one parameter: it must modify the cosmic history in the correct epochs and in the correct directions.

\subsection{The Hubble Tension as a Sound-Horizon Problem}

The most mature and statistically sharp anomaly is the discrepancy in the Hubble constant. Within flat $\Lambda$CDM, early-universe data calibrated by the cosmic microwave background (CMB), especially when combined with external information such as lensing and baryon-acoustic-oscillation measurements, favor a lower value of $H_{0}$ than direct late-time determinations based on the local distance ladder \cite{Planck2018,ACTDR6LCDM2025,Riess2022}. Alternative late-time methods, such as tip-of-the-red-giant-branch measurements and strong-lensing time delays, typically lie between the CMB-preferred and SH0ES values, but still tend to remain above the standard early-time anchor \cite{Freedman2025,Birrer2025,Verde2019,KnoxMillea2020}.

The CMB does not measure $H_{0}$ directly. What it measures extremely well is the angular acoustic scale,

\begin{equation}
\theta_{*} = \frac{r_{s}(z_{*})}{D_{A}(z_{*})},
\end{equation}
where $r_{s}(z_{*})$ is the comoving sound horizon at last scattering and $D_{A}(z_{*})$ is the angular-diameter distance to the last-scattering surface. The sound horizon is \cite{HuSugiyama1995,EisensteinHu1998}
\begin{equation}
r_{s}(z_{*}) = \int_{z_{*}}^{\infty} \frac{c_{s}(z)}{H(z)}\,dz,
\qquad
c_{s}(z)=\frac{1}{\sqrt{3\left[1+R(z)\right]}}.
\end{equation}

Thus,
 the early-time inference of $H_{0}$ is tied to the pre-recombination expansion history through $r_{s}$. If one increases $H(z)$ before recombination while leaving the rest of the fit under control, then $r_{s}$ decreases. Since $\theta_{*}$ is tightly measured, a smaller $r_{s}$ must be compensated mainly by a smaller distance to last scattering, which generally pushes the inferred $H_{0}$ upward.

\textls[-15]{This is why the Hubble tension is, in practice, largely a sound-horizon tension \cite{Verde2019,KnoxMillea2020,DiValentino2021,Poulin2019,PoulinSmithKarwal2023}.} The question is not merely whether one can add energy density to the early universe, but whether one can do so in a localized and sufficiently controlled way: large enough to reduce $r_{s}$, early enough to affect the acoustic scale, and transient enough not to spoil the rest of the thermal history. In that sense, the tension is highly constraining. It demands a modification near the pre-recombination era, but it also demands that this modification disappear quickly afterward.

\subsection{The Dark-Energy Tension as a History-of-Acceleration Problem}

A second, qualitatively different issue has emerged from recent analyses that combine baryon-acoustic-oscillation information with CMB data and, in some combinations, supernova distances \cite{DESI2025DR2,ACTDR6Extended2025,Brout2022PantheonPlus,Abbott2024DESSN5YR}. In the common Chevallier--Polarski--Linder (CPL) description \cite{ChevallierPolarski2001,Linder2003},
\begin{equation}
w(a) = w_{0} + w_{a}(1-a),
\end{equation}
the preferred region is often displaced from the cosmological-constant point $(w_{0},w_{a}) = (-1,0)$ toward the quadrant
\begin{equation}
w_{0} > -1,
\qquad
w_{a} < 0.
\end{equation}

This is the sign pattern associated with a dark-energy component that is close to a cosmological constant today but was more important in the past than a pure $\Lambda$ would be \cite{CaldwellLinder2005}.

 Here, the issue is not primarily the sound horizon, but the late-time expansion history. A component with $w>-1$ redshifts more slowly than matter but faster than a true cosmological constant, so for a fixed present-day dark-energy density, it contributes a larger fraction of the energy budget at intermediate redshift. In other words, it modifies when acceleration becomes important and how rapidly the universe interpolates between matter domination and dark-energy domination.

It is important to state the status of this anomaly carefully. The evidence for evolving dark energy is not as settled as the existence of the $H_{0}$ discrepancy, and its significance depends on the exact dataset combination and modeling assumptions \cite{DESI2025DR2,ACTDR6Extended2025,Brout2022PantheonPlus,Abbott2024DESSN5YR}. Nevertheless, the repeated appearance of the same preferred quadrant is striking enough to motivate model building. For the purposes of the present work, the relevant lesson is operational: if one wants a single mechanism to address current hints in the data, then its late-time behavior should naturally favor $w_{0}>-1$ and $w_{a}<0$, rather than requiring a finely tuned trajectory that crosses the phantom divide or oscillates artificially.

\subsection{The $S_{8}$ Tension as a Growth-of-Structure Problem}

The third target concerns the amplitude of late-time matter clustering. It is commonly summarized by the parameter
\begin{equation}
S_{8} \equiv \sigma_{8}\left(\frac{\Omega_{m0}}{0.3}\right)^{1/2},
\end{equation}
which captures the degeneracy direction to which weak-lensing and low-redshift large-scale-structure surveys are especially sensitive. The broad empirical pattern is that several late-time probes have tended to prefer somewhat smaller clustering amplitudes than the values inferred from CMB-calibrated $\Lambda$CDM fits, although the size of the discrepancy is modest and clearly probe-dependent \cite{DESY6Shear2026,KiDSLegacy2025,Planck2018,ACTDR6LCDM2025}.

This tension is physically different from both the $H_{0}$ and $w_{0}$–$w_{a}$ issues. It is not mainly about distances, but about the efficiency with which matter perturbations grow once the primordial fluctuations have been laid down. In general relativity with smooth dark energy, the linear growth factor $D(a)$ obeys \cite{Peebles1980,LinderGrowth2005}
\begin{equation}
D'' + \left(2 + \frac{H'}{H}\right) D' - \frac{3}{2}\Omega_{m}(a)D = 0,
\end{equation}
where primes denote derivatives with respect to $\ln a$. This equation shows transparently that growth is controlled by two competing effects: the matter fraction drives clustering, while cosmic expansion produces friction. Any mechanism that makes $H(a)$ larger than in $\Lambda$CDM during the era of structure formation, or that lowers $\Omega_{m}(a)$, suppresses the accumulated growth.

That is why late-time quintessence-like behavior tends to lower $S_{8}$ \cite{CaldwellLinder2005,LinderGrowth2005}. If dark energy is already slightly more important at intermediate redshift than in a pure cosmological-constant model, then the universe expands faster during the growth era and matter has less time to cluster efficiently. The resulting suppression is usually modest, which is appropriate here: the $S_{8}$ anomaly is not a dramatic failure of the standard model, but rather a persistent low-level preference in some datasets. A viable mechanism should therefore reduce growth without overcorrecting it.

\subsection{Why These Are Three Separate Targets}

The three tensions are linked observationally only in the weak sense that they all involve comparisons between early- and late-time inferences. Dynamically, however, they point to different cosmic epochs and different observables:
\begin{align}
H_{0}\ \text{tension} &\quad \Longrightarrow \quad \text{pre-recombination } H(z)\ \text{and } r_{s}, \\
(w_{0},w_{a})\ \text{trend} &\quad \Longrightarrow \quad \text{late-time expansion history}, \\
S_{8}\ \text{tension} &\quad \Longrightarrow \quad \text{late-time linear growth}.
\end{align}

This distinction matters for model building. An early-time modification can reduce the sound horizon and help with $H_{0}$, but by itself, it does not guarantee the preferred late-time dark-energy behavior. Conversely, a late-time scalar field can generate $w_{0}>-1$, $w_{a}<0$, and weaker growth, but it cannot on its own change the pre-recombination sound horizon enough to move the CMB-inferred Hubble constant. The challenge is therefore not to solve one of these problems in isolation, but to construct a single framework that acts in the right way at both early and late times.

This is the perspective adopted in the present paper. I seek one dynamical degree of freedom whose evolution naturally breaks into three stages: a transient pre-recombination energy injection that lowers $r_{s}$, a rapid post-event dilution so that standard cosmology is recovered soon afterward, and a late-time quintessence phase that produces $w_{0}>-1$, $w_{a}<0$, and weaker structure growth. The review above shows why each of these ingredients is required. The next section shows how one smooth scalar potential can realize all of them in a single continuous history.

\section{Exact One-Field Dynamics and Observational Calibration}
\label{sec:resolution}

The model is defined by the Einstein-frame action
\begin{equation}
\begin{aligned}
S={}&\int d^4x\sqrt{-g}\left[\frac{M_{\rm Pl}^2}{2}R
-\frac12g^{\mu\nu}\partial_\mu\phi\partial_\nu\phi-V(\phi)\right]\\
&+S_m[g_{\mu\nu},\Psi_m].
\end{aligned}
\label{eq:action}
\end{equation}
with a single canonical field and
\begin{equation}
V(\phi)=V_0 e^{-\lambda\phi/M_{\rm Pl}}
\left[1+A\exp\!\left(-\frac{(\phi-\phi_c)^2}{2\sigma^2}\right)\right].
\label{eq:potential}
\end{equation}

\begin{figure}[H]

\begin{adjustwidth}{-\extralength}{0cm}
\centering
 \includegraphics[width=1.2\textwidth]{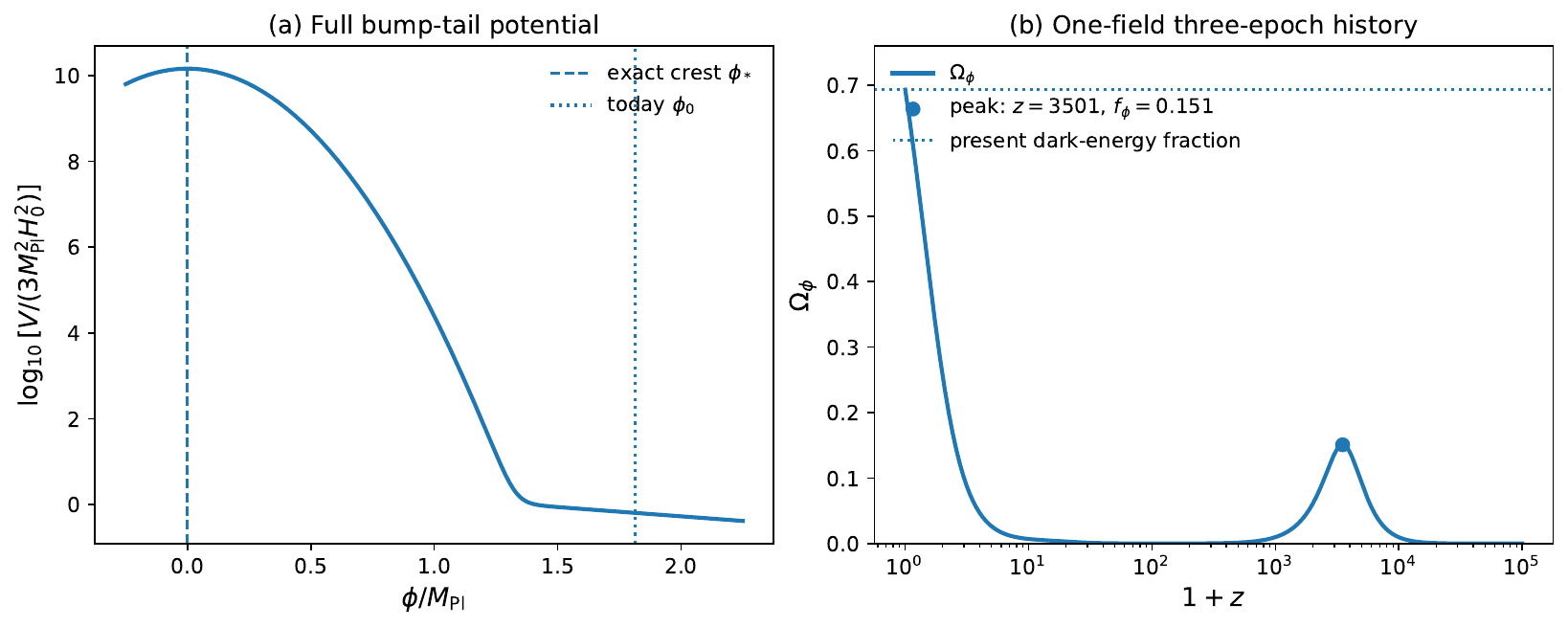}
\end{adjustwidth}
    \caption{Full-potential
 benchmark. \textbf{(a)}:
 the exact bump--tail potential, with the exact crest position $\phi_*$ and present field value, $\phi_0$, marked. \textbf{(b)}:
  the scalar energy fraction obtained by integrating the same potential from the radiation era to today. The localized feature produces a transient peak, $\Omega_{\phi,\rm pk}=0.151$, at $z_{\rm pk}\simeq3501$; after rapid dilution, the field re-emerges on the exponential tail and supplies the present dark-energy density.}
    \label{fig:potential_history}
\end{figure}

No modification of gravity, non-minimal coupling, or dark-sector interaction is introduced. Figure \ref{fig:potential_history} shows both the full potential and the numerically evolved scalar fraction for the benchmark specified in Table \ref{tab:benchmark}; importantly, the early and late curves are not separate proxy cosmologies but two epochs of the same solution of Equations \eqref{eq:friedmann_full} and \eqref{eq:kg_full}.

\vspace{-6pt}
\begin{table}[H]
\caption{Numerical
 benchmark used in Figures \ref{fig:potential_history}--\ref{fig:growth_prediction}. Quantities labeled ``derived'' are outputs of the same full-potential integration, not independent inputs.}
\label{tab:benchmark}
\begin{tabularx}{\textwidth}{l C}
\toprule
\textbf{Quantity} & \textbf{Value}\\
\midrule
$H_0$ [km s$^{-1}$ Mpc$^{-1}$] & 70.0\\
$\Omega_{m0}$ & 0.306\\
$\omega_b\equiv\Omega_b h^2$ & 0.02237\\
high-$z$ normalization $\sigma_{8,\rm ref}$ & 0.811\\
$\lambda$ & 0.9898\\
$A$ & $3.85\times10^9$\\
$\sigma/M_{\rm Pl}$ & 0.194\\
$\phi_c/M_{\rm Pl}$ ($\phi_*=0$ convention) & 0.03725\\
$V_0/(3M_{\rm Pl}^2H_0^2)$ & 3.8464\\
$\phi_i/M_{\rm Pl}$ at $N_i=-12$ & 0.1358\\
\midrule
$z_{\rm pk}$ (derived) & 3501\\
$\Omega_{\phi,\rm pk}$ (derived) & 0.151\\
$r_d$ [Mpc] (derived) & 141.1\\
$w_0$ (derived) & $-0.844$\\
$w_a$ local CPL projection (derived) & $-0.176$\\
$S_8$ (derived) & 0.794\\
$\chi^2_{\rm BAO}$, 13 DESI DR2 observables & 9.48\\
\bottomrule
\end{tabularx}
\end{table}

\vspace{-6pt}
\begin{figure}[H]
    \includegraphics[width=0.96\columnwidth]{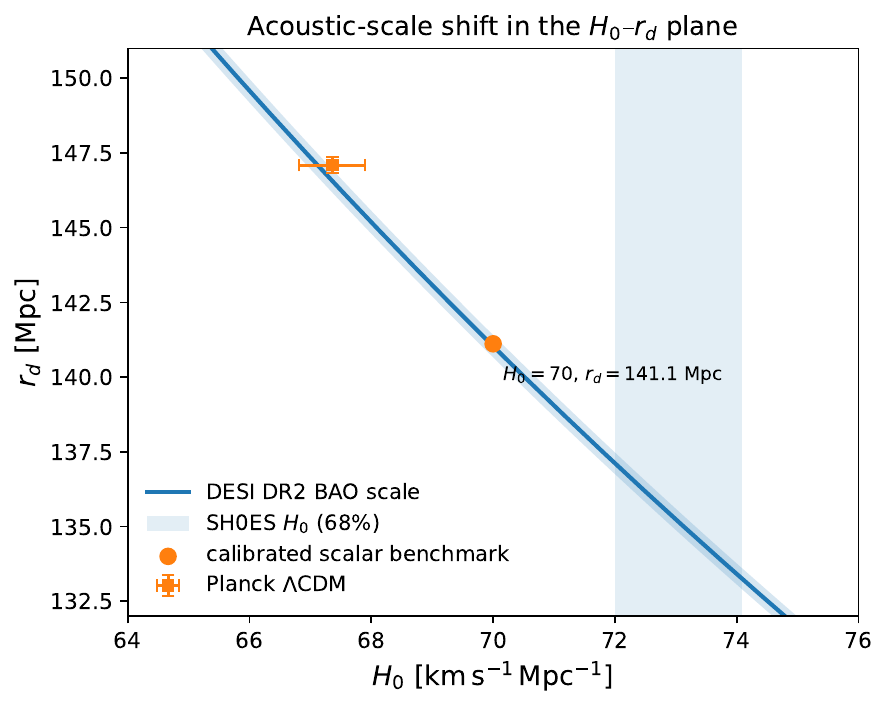}
    \caption{Quantitative
 $H_0$--$r_d$ comparison. The DESI DR2 band is obtained from the BAO likelihood for the scale $c/(H_0r_d)$; the vertical band is the SH0ES $68\%$ interval, and the square shows the Planck $\Lambda$CDM determination. The calibrated full scalar benchmark (circle) gives $(H_0,r_d)=(70.0,141.1\,\mathrm{Mpc})$, moving coherently toward larger $H_0$ and smaller $r_d$ while remaining on the BAO distance scale.}
    \label{fig:h0rd_quantitative}
\end{figure}

\vspace{-6pt}

\begin{figure}[H]

\begin{adjustwidth}{-\extralength}{0cm}
\centering
    \includegraphics[width=1.2\textwidth]{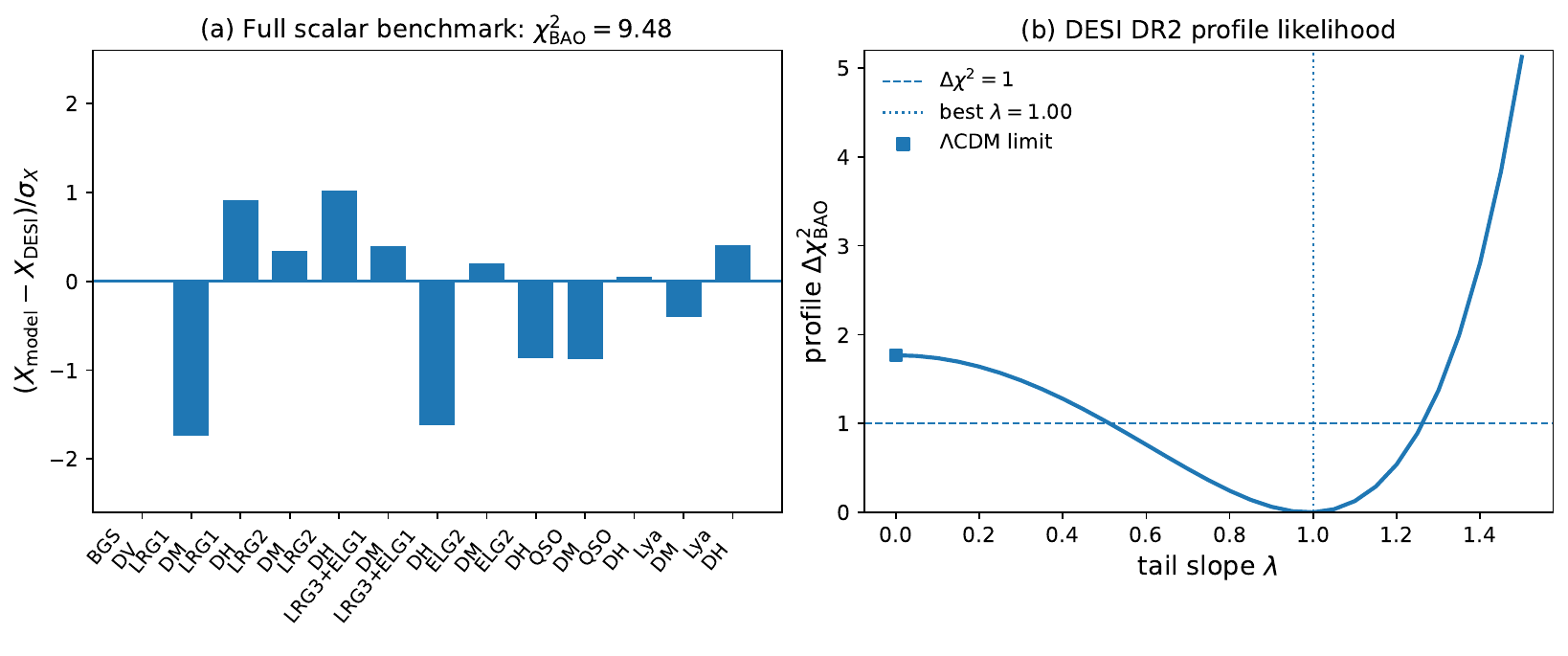}
\end{adjustwidth}
    \caption{DESI DR2 quantitative test. Left: pulls of all 13 BAO observables for the full scalar benchmark; the reported within-pair correlations are included when computing $\chi^2_{\rm BAO}=9.48$. Right: profile likelihood for the exact scalar tail after profiling over $\Omega_{m0}$ and $c/(H_0r_d)$. The minimum occurs at $\lambda\simeq1.00$; the horizontal line marks $\Delta\chi^2=1$, and the $\lambda=0$ limit is shown explicitly.}
    \label{fig:desi_likelihood}
\end{figure}
\unskip

\begin{figure}[H]
    \includegraphics[width=0.96\columnwidth]{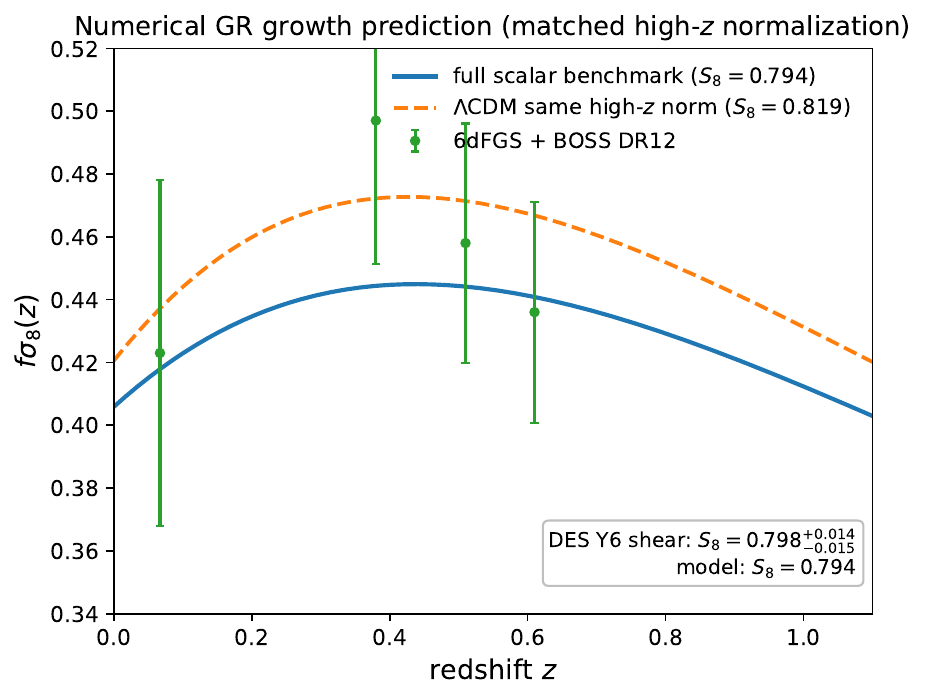}
    \caption{Numerical GR growth prediction from the full benchmark, conditional on the matched high-redshift normalization. The solid curve is obtained by integrating Equation \eqref{eq:growth_exact} on the same scalar background used in Figures \ref{fig:potential_history}--\ref{fig:desi_likelihood}; the dashed curve is a matched $\Lambda$CDM background with the same high-redshift normalization. The points are 6dFGS and BOSS DR12 $f\sigma_8$ measurements. The scalar trajectory produces a $3.0\%$ reduction in the present growth factor and predicts $S_8=0.794$, consistent with DES Y6 cosmic shear.}
    \label{fig:growth_prediction}
\end{figure}

\subsection{Physical Origin and Effective-Theory Interpretation of the Potential}

The two factors in Equation \eqref{eq:potential} have a standard effective-field-theory interpretation. Exponential scalar potentials arise naturally for canonically normalized moduli after dimensional reduction and occur in supergravity/M-theory constructions, in addition to their long history as scaling and quintessence potentials \cite{Townsend2001,CopelandLiddleWands1998,CopelandSamiTsujikawa2006}. The localized factor is most usefully viewed as the minimal analytic representative of a smooth localized correction to that runaway direction. For any positive localized contribution centered near $\phi_c$, its behavior over the field interval sampled during release is controlled first by its height, position, and curvature; a Gaussian packages precisely these three quantities while keeping the complete potential differentiable. The cosmological mechanism therefore belongs to a broader ``localized feature + runaway tail'' universality class and does not rely on the Gaussian being a unique ultraviolet completion.

There is also less independent parameter freedom than the five symbols in \linebreak  Equation \eqref{eq:potential} might suggest. A constant field translation can be absorbed into $V_0$ and $\phi_c$, so one field-space origin is redundant. I use this freedom to set the exact crest to $\phi_*=0$. The present flatness condition fixes the overall normalization $V_0$ once $H_0$, $\Omega_{m0}$ and the trajectory are specified, while the exact extremum condition below fixes $\phi_c$ for chosen $(\lambda,A,\sigma)$. The physically relevant quantities are consequently combinations such as the tail slope, $\lambda$, the feature width, $\sigma/M_{\rm Pl}$, the feature-to-tail hierarchy, $A$, and the initial displacement from the crest. In particular, the large numerical value of $A$ in Table \ref{tab:benchmark} is the dimensionless representation of the many-orders-of-magnitude energy-density hierarchy between a feature active near $z\sim10^3$ and the dark-energy tail today, rather than five unrelated knobs independently adjusted at late times.

\subsection{Exact Extremum and Release Condition}

To avoid a bump-dominated approximation, define
\begin{equation}
\begin{aligned}
\Delta&\equiv\phi-\phi_c,\qquad
G(\phi)\equiv A e^{-\Delta^2/(2\sigma^2)},\\
q(\phi)&\equiv\frac{G}{1+G}.
\end{aligned}
\end{equation}

Differentiating the complete potential gives
\begin{equation}
\frac{V_{,\phi}}{V}=-\frac{\lambda}{M_{\rm Pl}}-\frac{\Delta}{\sigma^2}q,
\label{eq:exact_d1}
\end{equation}
and
\begin{equation}
\frac{V_{,\phi\phi}}{V}=
\left(\frac{V_{,\phi}}{V}\right)^2
-\frac{q}{\sigma^2}
+\frac{\Delta^2}{\sigma^4}q(1-q).
\label{eq:exact_d2}
\end{equation}

Let $y\equiv(\phi_c-\phi_*)/\sigma>0$ and $\beta\equiv\lambda\sigma/M_{\rm Pl}$. The exact stationary-point equation $V_{,\phi}(\phi_*)=0$ is then
\begin{equation}
\;\beta=y_*\,q_*\;,\qquad
q_*=\frac{A e^{-y_*^2/2}}{1+A e^{-y_*^2/2}}.
\label{eq:exact_extremum}
\end{equation}

At that point,
\begin{equation}
\;\frac{V_{,\phi\phi}(\phi_*)}{V(\phi_*)
=\frac{q_*}{\sigma^2}\left[-1+y_*^2(1-q_*)\right]\;}.
\label{eq:exact_curvature}
\end{equation}

Thus, the stationary point is a local maximum when $y_*^2(1-q_*)<1$. Equations \eqref{eq:exact_extremum} and \eqref{eq:exact_curvature} are exact for the full exponential-times-Gaussian potential and replace the earlier estimate $V_{,\phi\phi}\simeq-V/\sigma^2$.

A useful analytic diagnostic follows by defining release through $3H_c\simeq\sqrt{-V_{,\phi\phi}(\phi_*)}$ and taking the temporarily frozen field to be potential dominated. The corresponding local estimate is
\begin{equation}
f_{\phi,c}\equiv\frac{V(\phi_*)}{3M_{\rm Pl}^2H_c^2}
\simeq
\frac{3\sigma^2/M_{\rm Pl}^2}
{q_*\left[1-y_*^2(1-q_*)\right]}.
\label{eq:fede_exact}
\end{equation}

The familiar $3\sigma^2/M_{\rm Pl}^2$ expression is recovered only in the controlled limit $A\gg1$, $q_*\to1$. For the benchmark, Equation \eqref{eq:fede_exact} gives $f_{\phi,c}\simeq0.113$, whereas the exact finite-displacement evolution reaches $\Omega_{\phi,\rm pk}=0.151$; the latter is the quantity used in all numerical figures.

\subsection{Full Background Evolution and Benchmark}

For a flat FLRW background, the exact equations integrated in this work are
\begin{equation}
H^2=\frac{1}{3M_{\rm Pl}^2}
\left(\rho_m+\rho_r+\frac12\dot\phi^2+V\right),
\label{eq:friedmann_full}
\end{equation}
\begin{equation}
\ddot\phi+3H\dot\phi+V_{,\phi}=0,
\label{eq:kg_full}
\end{equation}
with standard conservation for matter and radiation. I integrate in $N=\ln a$ from $N_i=-12$, use $\dot\phi_i=0$, and impose the present flatness condition by shooting on $V_0$. The initial displacement is stated explicitly in Table \ref{tab:benchmark}; it is therefore not hidden in the numerical construction.

The benchmark is deliberately chosen as a single reproducible point rather than a collection of visually selected curves. Its standard and scalar parameters, together with the most relevant derived observables, are listed in Table \ref{tab:benchmark}. The feature peaks shortly before recombination and is followed by a fast-roll phase. When the kinetic term dominates, $w_\phi\to1$ and $\rho_\phi\propto a^{-6}$, so the early contribution automatically becomes negligible after release. At late times, the same field reaches the exponential tail, where it thaws into a smooth quintessence component.

\subsection{Sound Horizon and the $H_0$--$r_d$ Direction}

The drag-epoch sound horizon is computed directly from the numerical background,
\begin{equation}
r_d=\int_{z_d}^{\infty}\frac{c_s(z)}{H(z)}\,dz,
\qquad
c_s=\frac{c}{\sqrt{3[1+R(z)]}},
\end{equation}
where $R=3\rho_b/(4\rho_\gamma)$, and I use $z_d=1059.94$ for the benchmark calibration. The transient scalar contribution raises $H(z)$ around the pre-recombination epoch and yields $r_d=141.1$ Mpc. Together with $H_0=70.0$ km s$^{-1}$ Mpc$^{-1}$, the dimensionless BAO scale is
\begin{equation}
\alpha\equiv\frac{c}{H_0r_d}=30.350,
\end{equation}
which is extremely close to the DESI DR2 BAO-only best-fit value $30.367$ obtained below. Figure \ref{fig:h0rd_quantitative} therefore replaces a qualitative ``preferred-direction'' plot by actual numerical values and $68\%$ uncertainty regions from Planck $\Lambda$CDM, SH0ES, and the DESI BAO scale \cite{Planck2018,Riess2022,DESI2025DR2}.

\subsection{DESI DR2 Likelihood and Observational Estimate of $\lambda$}

I now quantify the BAO statement rather than judging the agreement by eye. I use the DESI DR2 measurements, $D_V/r_d$, for BGS and the correlated $(D_M/r_d,D_H/r_d)$ pairs for LRG1, LRG2, LRG3+ELG1, ELG2, QSO and Ly$\alpha$, with the reported $1\sigma$ uncertainties and within-pair $D_M$--$D_H$ correlation coefficients \cite{DESI2025DR2}. For a residual vector, $\Delta\mathbf d$, with the block covariance, $\mathbf C$, constructed from those quoted uncertainties and correlations, the statistic is
\begin{equation}
\chi^2_{\rm BAO}=\Delta\mathbf d^{\rm T}\mathbf C^{-1}\Delta\mathbf d.
\label{eq:bao_chi2}
\end{equation}
The full benchmark in Table \ref{tab:benchmark} gives $\chi^2_{\rm BAO}=9.48$ for 13 measured quantities. This is the statistic quoted in Figure \ref{fig:desi_likelihood}; no visual rescaling of the data is used.

The tail slope can be constrained particularly cleanly because it controls the late-time departure from a cosmological constant. To isolate this dependence I profile the exact exponential-tail scalar background over $\Omega_{m0}$ and the BAO scale, $\alpha=c/(H_0r_d)$, at each $\lambda$. The result is
\begin{equation}
\lambda_{\rm bf}=1.00,
\qquad
0.51<\lambda<1.26\quad(\Delta\chi^2\leq1),
\label{eq:lambda_constraint}
\end{equation}
with $\chi^2_{\rm min}=8.78$, while the $\lambda=0$ (cosmological-constant tail) limit gives $\chi^2=10.55$. Thus, DESI BAO alone already selects the physically relevant order of magnitude $\lambda=\mathcal O(1)$ for this scalar trajectory. Equation \eqref{eq:lambda_constraint} is a profile-likelihood calibration, not a substitute for a joint Bayesian posterior including the full CMB spectra; its purpose is to answer directly which tail slopes are favored by the BAO distances and to provide an observationally anchored benchmark for the full model.

\subsection{Dark-Energy Evolution Without Assuming CPL Dynamics}

The model does not assume the CPL equation of state. At every time, I calculate
\begin{equation}
w_\phi(a)=\frac{\dot\phi^2/2-V}{\dot\phi^2/2+V}
\label{eq:wphi_exact}
\end{equation}
directly from the Klein--Gordon solution. CPL is used only as a two-number observational compression near $a=1$,
\begin{equation}
w_0\equiv w_\phi(1),\qquad
w_a\equiv-\left.\frac{dw_\phi}{d\ln a}\right|_{a=1},
\label{eq:cpl_projection}
\end{equation}
which gives $(w_0,w_a)=(-0.844,-0.176)$ for the benchmark. Another phenomenological basis---constant $w$, redshift bins, a pivot expansion, or principal components---would merely project the same numerical function, $w_\phi(a)$, onto different basis functions; it would not change the dynamics of Equation \eqref{eq:kg_full}. Distances and growth in our likelihood calculation are therefore integrated from $H(z)$ itself, not reconstructed from CPL.

This distinction also yields a sharp physical prediction. For a canonical minimally coupled field with positive energy density,
\begin{equation}
1+w_\phi=\frac{\dot\phi^2}{\rho_\phi}\ge0,
\end{equation}
so the trajectory is non-phantom and cannot cross $w=-1$. The model consequently occupies the quintessence sector $w_0>-1$ and, for the present thawing trajectory, $w_a<0$, the same sign direction highlighted by recent DESI combinations with CMB and supernova data \cite{DESI2025DR2,Brout2022PantheonPlus,CaldwellLinder2005}. This is a falsifiable restriction of the scalar mechanism rather than an assumption imported from the CPL parametrization.

\subsection{Perturbations, Stability, and a Genuine Growth Prediction}

Because the field is canonical, the quadratic kinetic term has the standard positive sign and the rest-frame scalar sound speed is
\begin{equation}
c_{s,\phi}^2=1.
\end{equation}
Hence, there is neither a ghost nor a gradient instability. In Newtonian gauge,
\begin{equation}
ds^2=-(1+2\Psi)dt^2+a^2(1-2\Phi)d\mathbf x^2,
\end{equation}
the Fourier-space scalar perturbation obeys the standard canonical equation \cite{Mukhanov2005,CopelandSamiTsujikawa2006}
\begin{equation}
\delta\ddot\phi+3H\delta\dot\phi+
\left(\frac{k^2}{a^2}+V_{,\phi\phi}\right)\delta\phi
=4\dot\phi\dot\Phi-2V_{,\phi}\Psi,
\label{eq:deltaphi}
\end{equation}
with $\Phi=\Psi$ in the absence of anisotropic stress from the scalar. Near the crest, $V_{,\phi\phi}<0$ by construction, which supplies the desired finite-duration hilltop release. This is a tachyonic curvature of the potential, not a pathology of the kinetic sector; once the field leaves the localized feature, the instability terminates dynamically.

For the late-time clustering test, $c_{s,\phi}^2=1$ makes scalar fluctuations smooth on the sub-horizon scales used by RSD and weak lensing. The matter growth factor therefore satisfies, to excellent accuracy,
\begin{equation}
D''+\left(2+\frac{H'}{H}\right)D'-\frac32\Omega_m(a)D=0,
\label{eq:growth_exact}
\end{equation}
where primes denote $d/d\ln a$ and $H(a)$ is the same numerical background used for the sound horizon and BAO calculations. I solve Equation \eqref{eq:growth_exact} from $z=50$ with the same high-redshift normalization as the matched Planck-normalized reference, $\sigma_{8,\rm ref}=0.811$. The full scalar trajectory gives
\begin{equation}
\begin{aligned}
\frac{D_\phi(1)}{D_{\Lambda\mathrm{CDM}}(1)}&=0.9696,\\
\sigma_8(0)&=0.786,\qquad S_8=0.794.
\end{aligned}
\label{eq:growth_numbers}
\end{equation}
The DES Year-6 cosmic-shear result, $S_8=0.798^{+0.014}_{-0.015}$, for its preferred NLA analysis, therefore lies essentially on the benchmark prediction \cite{DESY6Shear2026}. Figure \ref{fig:growth_prediction} additionally compares the predicted $f\sigma_8(z)$ curve with 6dFGS and BOSS DR12 RSD measurements \cite{Beutler2012,Alam2017}.
 The RSD points are used as an external growth cross-check; the plotted scalar line is no longer an effective proxy.

\subsection{What Is Constrained Now and What a Full-Spectrum Test Must Add}

The analysis intentionally separates quantities that can be tested with the observables evaluated here from those that require a Boltzmann-code likelihood. The DESI-compressed BAO likelihood directly constrains the late-time slope and distance scale; the sound-horizon integral constrains the early feature through $r_d$; DES Y6 and RSD test the associated growth suppression; and the local $H_0$ measurement tests whether the early-time reduction in $r_d$ moves the inverse-distance-ladder calibration in the required direction. CMB temperature/polarization and CMB lensing are additionally crucial because the early scalar component changes the pre-recombination gravitational potentials and therefore cannot be reduced to $r_d$ alone. A statistically definitive claim of ``resolution'' must consequently fit TT/TE/EE and lensing with Equation \eqref{eq:deltaphi} implemented in a Boltzmann solver, jointly with BAO, Pantheon+, RSD, weak lensing, and a local-$H_0$ likelihood.

The revised claim is correspondingly precise: the present calculation establishes that the same exact canonical trajectory can (i) generate a localized early fraction and reduce $r_d$, (ii) fit the DESI DR2 BAO distance ratios quantitatively with an order-unity, observationally profiled $\lambda$, and (iii) suppress late growth to the DES-Y6 $S_8$ range, while remaining perturbatively well posed. This is the nontrivial prerequisite that a full joint likelihood must test rather than assume.

\section{Conclusions}
\label{sec:conclusions}

I have shown that a single canonical scalar field in standard general relativity can realize a quantitatively calibrated sequence of early energy injection, rapid post-release dilution, and late-time quintessence within one smooth potential. The revised analysis goes beyond the original qualitative construction in three decisive respects. First, the position and curvature of the localized maximum are obtained exactly from the complete potential, Equations \eqref{eq:exact_extremum} and \eqref{eq:exact_curvature}. Second, all principal figures now follow from one full numerical trajectory rather than separate effective proxies. Third, the comparison with DESI DR2 uses the published compressed BAO errors and correlations and yields an explicit likelihood profile for the tail slope.

The benchmark makes the correlated phenomenology concrete. It produces $\Omega_{\phi,\rm pk}=0.151$ at $z\simeq3501$, lowers the drag sound horizon to $r_d=141.1$ Mpc, and supports $H_0=70.0$ km s$^{-1}$ Mpc$^{-1}$ along the DESI BAO distance scale. The BAO profile gives $\lambda_{\rm bf}\simeq1.00$ with an approximate $68\%$ profile range $0.51<\lambda<1.26$. At late times the exact scalar history has a non-phantom thawing behavior whose local CPL projection is $(w_0,w_a)=(-0.844,-0.176)$. Solving the GR growth equation on that same background suppresses the present growth factor by $3.0\%$ relative to the matched $\Lambda$CDM case and gives $S_8=0.794$, essentially coincident with the DES Year-6 cosmic-shear determination. The field is canonically stable, with $c_s^2=1$; the negative curvature at the early crest is the controlled dynamical trigger for release rather than a ghost or gradient instability.

The importance of the construction is therefore its economy and correlation structure. The sound-horizon shift and the late growth suppression are not produced by unrelated sectors or by modifying gravity: both are consequences of one scalar trajectory whose parameters have direct observational roles. The exponential slope is measured by the late expansion history, the localized feature controls the early fraction and $r_d$, and the same background fixes the growth history. This makes the model considerably more restrictive than an arbitrary independent early-dark-energy component plus a separately parameterized late dark-energy sector.

A full TT/TE/EE+lensing implementation remains the decisive precision test, and the title and claims have been formulated accordingly: the present work establishes a unified, exact, and quantitatively data-calibrated mechanism rather than asserting a completed global model-selection result. The next step is therefore sharply defined---implement Equations \eqref{eq:kg_full} and \eqref{eq:deltaphi} in a Boltzmann code and perform a joint CMB+lensing+DESI+Pantheon++RSD+weak-lensing+$H_0$ analysis. Because the mechanism already yields a single trajectory with the required early and late signatures, that calculation is a direct falsifiability test of the model rather than a search for whether the desired phenomenology can be engineered at all.

\vspace{+6pt}

\funding{This research received no external funding.
}

\dataavailability{No new observational or experimental data were generated in this study. The observational data used in the analysis are publicly available from the sources cited in the article. The numerical results can be reproduced from the equations, benchmark parameters, and observational inputs described in the article.
}

\conflictsofinterest{The author declares no conflicts of interest.}

\begin{adjustwidth}{-\extralength}{0cm}

\reftitle{References}

\PublishersNote{}
\end{adjustwidth}
\end{document}